\documentclass[twocolumn]{aastex631}

\usepackage{xcolor}
\usepackage{verbatim}
\usepackage{amsmath}

\newcommand{\sersic}{S\'{e}rsic}

\begin{document}

\shorttitle{Size Growth on Short Timescales of Star-Forming Galaxies}


\title{Size Growth on Short Timescales of Star-Forming Galaxies: Insights from Size Variation with Rest-Frame Wavelength with JADES}

\author{Cheng Jia}\thanks{E-mail:  jc123@mail.ustc.edu.cn}
\affiliation{Department of Astronomy, University of Science and Technology of China, Hefei 230026, China}
\affiliation{School of Astronomy and Space Science, University of Science and Technology of China, Hefei 230026, China}

\author[0000-0003-1588-9394]{Enci Wang}\thanks{E-mail: ecwang16@ustc.edu.cn}
\affiliation{Department of Astronomy, University of Science and Technology of China, Hefei 230026, China}
\affiliation{School of Astronomy and Space Science, University of Science and Technology of China, Hefei 230026, China}

\author{Huiyuan Wang}
\affiliation{Department of Astronomy, University of Science and Technology of China, Hefei 230026, China}
\affiliation{School of Astronomy and Space Science, University of Science and Technology of China, Hefei 230026, China}

\author[0000-0002-1253-2763]{Hui Li}
\affiliation{Department of Astronomy, Tsinghua University, Haidian DS 100084, Beijing, China}

\author{Yao Yao}
\affiliation{Department of Astronomy, University of Science and Technology of China, Hefei 230026, China}
\affiliation{School of Astronomy and Space Science, University of Science and Technology of China, Hefei 230026, China}

\author{Jie Song}
\affiliation{Department of Astronomy, University of Science and Technology of China, Hefei 230026, China}
\affiliation{School of Astronomy and Space Science, University of Science and Technology of China, Hefei 230026, China}

\author{Hongxin Zhang}
\affiliation{Department of Astronomy, University of Science and Technology of China, Hefei 230026, China}
\affiliation{School of Astronomy and Space Science, University of Science and Technology of China, Hefei 230026, China}

\author{Yu Rong}
\affiliation{Department of Astronomy, University of Science and Technology of China, Hefei 230026, China}
\affiliation{School of Astronomy and Space Science, University of Science and Technology of China, Hefei 230026, China}

\author{Yangyao Chen}
\affiliation{Department of Astronomy, University of Science and Technology of China, Hefei 230026, China}
\affiliation{School of Astronomy and Space Science, University of Science and Technology of China, Hefei 230026, China}

\author{Haoran Yu}
\affiliation{Department of Astronomy, University of Science and Technology of China, Hefei 230026, China}
\affiliation{School of Astronomy and Space Science, University of Science and Technology of China, Hefei 230026, China}

\author{Zeyu Chen}
\affiliation{Department of Astronomy, University of Science and Technology of China, Hefei 230026, China}
\affiliation{School of Astronomy and Space Science, University of Science and Technology of China, Hefei 230026, China}

\author{Haixin Li}
\affiliation{Department of Astronomy, University of Science and Technology of China, Hefei 230026, China}
\affiliation{School of Astronomy and Space Science, University of Science and Technology of China, Hefei 230026, China}

\author{Chengyu Ma}
\affiliation{Department of Astronomy, University of Science and Technology of China, Hefei 230026, China}
\affiliation{School of Astronomy and Space Science, University of Science and Technology of China, Hefei 230026, China}

\author{Xu Kong}
\affiliation{Department of Astronomy, University of Science and Technology of China, Hefei 230026, China}
\affiliation{School of Astronomy and Space Science, University of Science and Technology of China, Hefei 230026, China}

\begin{abstract}

We investigate size variation with rest-frame wavelength for 
star-forming galaxies 
based on the second JWST Advanced Deep Extragalactic Survey data release.   Star-forming galaxies are typically smaller at longer wavelength from UV-to-NIR at $z<3.5$, especially for more massive galaxies, indicating the inside-out assembly with in-situ star formation if ignoring dust attenuation.  The size variation with wavelength shows strong dependence on stellar mass, and shows little or no dependence on redshift, specific star formation rate and galaxy environment.  This suggests that the size growth of star-forming galaxies is a self-regulated process primarily governed by stellar mass.  We model size as a function of both mass and redshift simultaneously, obtaining $R_{\rm e} \propto M_*^{0.23} (1+z)^{-1.04}$ at a wavelength of 0.45 ${\mu \mathrm{m}}$, and $R_{\rm e} \propto M_*^{0.20} (1+z)^{-1.08}$ at 1.0 ${\mu \mathrm{m}}$.  Based on this size evolution and the star formation main sequence from the literature, we obtain the locus of typical size growth for individual galaxies of different masses on the mass-size plane.   The moving trend of galaxies on the mass-size plane, which indicates the slopes of their locus, strongly correlates with the size ratio between 0.45 ${\mu \mathrm{m}}$ and 1.0 ${\mu \mathrm{m}}$, supporting the idea that the size variation with wavelength provides important information on size growth of galaxies  on short timescales.

\end{abstract}


\section{introduction}\label{sec:intro}

Understanding the relationship between mass and size in galaxies, particularly in the context of quiescent and star-forming galaxies, is crucial for unraveling the complexities of galaxy evolution \citep[e.g.][]{2003MNRAS.343..978S, 2014ApJ...788...28V}. The evolution of galaxy size with cosmic time offers insights into their assembly history and the interplay with their dark matter halos.


Numerous studies have delved into the complexities of the mass-size relation, revealing intriguing patterns and correlations across different galaxy populations \citep[e.g.][]{2003MNRAS.343..978S, 2014ApJ...788...28V, 2019ApJ...880...57M, 2019MNRAS.489.4135D, 2021MNRAS.506..928N, Ferguson_2004, Trujillo_2006, buitrago2008extremely, Carollo_2013, 2013ApJ...777..155O, 2015MNRAS.447.2603L, 2017ApJ...839...71F, 2021MNRAS.501.1028Y, 2024MNRAS.527.6110O, 2024ApJ...962..176W}. 
Early studies by \cite{2003MNRAS.343..978S}, based on the Sloan Digital Sky Survey \citep[SDSS;][]{2000AJ....120.1579Y} data,  highlighted the distinct dependencies between size and stellar mass in early and late-type galaxies. Late-type galaxies, characterized by ongoing star formation and prominent spiral arms, tend to exhibit larger sizes compared to early-type galaxies at a given mass, which are typically more spheroidal and feature older stellar populations.  Instead of separating galaxies by their morphological type, later studies of mass-size relation tend to divide the galaxies according to their star formation activities: star-forming galaxies and quiescent galaxies \citep{2014ApJ...788...28V}.   

Quiescent galaxies tend to be spheroidal in structure and show smaller half-light radii compared to star-forming galaxies of similar mass and redshift \citep[][]{2014ApJ...788...28V, 2019ApJ...880...57M}. This size difference is thought to be influenced by both ``nurture," such as post-quenching fading of stellar populations \citep{2015Sci...348..314T, Tacchella_2016, 2016ApJ...818..180C, Lilly_2016}, and ``nature," possibly reflecting different formation processes for spheroids and disks \citep{2013ApJ...776...63F, 2016ApJ...827L..32B, Wang-18}. 
On the other hand, star-forming galaxies exhibit a different trajectory on the mass-size plane. Their evolution, particularly in terms of size, is intricately linked to ongoing star formation processes and gas accretion from their circumgalactic medium. Studies have shown a correlation between stellar mass, effective radius, and the quenching process, especially for galaxies with stellar masses above a certain threshold \citep{van_der_Wel_2009, 2013ApJ...776...63F, Omand_2014, 2016ApJ...827L..32B, 2020ApJ...897..102C}. This suggests a nuanced interplay between internal mechanisms (e.g., stellar feedback, active galactic nuclei feedback) and environmental factors in driving the growth of galaxies and quenching \citep{2010ApJ...721..193P, 2017ApJ...846..139M, 2018ApJ...860..102W, 2020ApJ...895...25W}.

Benefits from advancements in observational capabilities, large surveys such as the 3D-HST survey \citep{2012ApJS..200...13B} and HST/WFC3 imaging from CANDELS \citep{2011ApJS..197...35G, 2011ApJS..197...36K}, have allowed for a deeper understanding of galaxy evolution \citep{2019ApJ...880...57M, 2021MNRAS.506..928N, 2024ApJ...962..176W}.  \cite{2014ApJ...788...28V} investigated galaxies at $z < 3$ and indicated that star-forming galaxies have a mass-size relation with $R_{\mathrm{eff}} \propto M_*^{0.21}$, and quiescent galaxies exhibit a much steeper relation with $R_{\mathrm{eff}} \propto M_*^{0.75}$. In addition, the size of star-forming galaxies decreases with redshift following $R_{\mathrm{eff}} \propto (1+z)^{-0.75}$, and the sizes of quiescent galaxies exhibit a much faster evolution following $R_{\mathrm{eff}} \propto (1+z)^{-1.48}$.  
Different from star-forming galaxies that can grow their sizes inside-out by in-situ star formation in the disks \citep{Wang-19, 2023ApJ...955...55W, du2024physical}, quiescent galaxies have increased by a factor of $\sim$3-5 in size since redshift of 2 via different mechanisms such as gas-poor minor mergers \citep{2007ApJ...671..285T, 2008ApJ...688..770F}.



Studies of galaxy size often face challenges due to redshift limitations and wavelength considerations. The HST is restricted to observing the rest-frame V band up to redshift of 2, while alternative measurements in the rest-frame ultraviolet have produced different results compared to optical studies \citep{2010ApJ...709L..21O, 2012ApJ...756L..12M, 2013ApJ...777..155O, 2015ApJS..219...15S, Holwerda_2015}. This is because galaxies show different sizes for different wavelengths \citep[e.g.][]{2014MNRAS.441.1340V, Yang_2022, Suess_2022, Costantin_2023, 2024MNRAS.527.6110O}. Therefore, consistent wavelength measurements are crucial for accurate size assessments.

Fortunately, the James Webb Space Telescope (JWST) has significantly improved our capabilities, extending observations in the rest-frame V band up to redshift of 7 with its NIRCam instrument, enhancing our understanding of galaxy sizes across cosmic epochs. It provides us a new window to learn about the size evolution of galaxies at near-infrared wavelength. 
Based on the Cosmic Evolution Early Release Survey \citep[CEERS;][]{2022ApJ...940L..55F, 2023ApJ...946L..13F, 2023ApJ...946L..12B}, \cite{2024ApJ...962..176W} and \cite{2024MNRAS.527.6110O} reviewed the mass-size relation of galaxies with considering the above effect, and obtained consistent results with many previous works at similar wavelengths.  With deep multi-band NIRCam images in GOODS-South from JADES, \cite{ji2024jades} explored the size evolution of a sample of 161 quiescent galaxies, at three different rest-frame wavelengths from UV-to-NIR.  All of these authors found that galaxies typically show larger size at shorter wavelength to a non-negligible extent, across the full range of wavelengths considered.

In this work, utilizing the second data release of JADES \citep{2023arXiv231012340E}, we aim to answer questions about how sizes vary with rest-frame wavelength of star-forming galaxies, and what controls the size variations. If ignoring the dust attenuation, the size variation with wavelength can reflect the distribution of newly formed stars with respect to older stellar populations. 
We aim to understand the distribution of stellar populations in galaxies with diverse physical properties.

This paper is structured as follows. In Section \ref{sec:data}, we briefly introduce the data we used and the sample selection. In Section \ref{sec:result}, we show the size variation with rest-frame wavelength, and explore the dependence on galaxy properties.   We construct a 2-dimensional fitting of size-mass-redshift relation for star-forming galaxies at two given wavelengths, 0.45 ${\mu \mathrm{m}}$ and 1.0 ${\mu \mathrm{m}}$, obtaining consistent results with previous works.  In Section \ref{sec:discuss}, we dedicate to explore the evolution locus of individual galaxies on the mass-size plan, and discuss the potential connection with size variation with wavelength.  We summarize our result in Section \ref{sec:summary}.

In this paper, we assume concordance flat $\mathrm{\Lambda CDM}$ cosmology with $\Omega_{m}=0.308$, $\Omega_{\Lambda}=0.691$, $H_0 = 67.74$ $\mathrm{km \cdot s^{-1} \cdot Mpc^{-1}}$ \citep{2016A&A...594A..13P}.

\section{data}\label{sec:data}
\subsection{JADES data and sample definition} \label{sec:2.1}



The data utilized in this study are from the JWST Advanced Deep Extragalactic Survey (JADES) data release version 2.0 \citep{2023arXiv230602465E, 2023ApJS..269...16R, 2023arXiv231012340E}, comprising observations in nine filters of the JWST NIRCam, specifically F090W,  F115W,  F150W,  F200W,  F277W, F335M,  F356W,  F410M, and F444W. 
It was obtained from the Mikulski Archive for Space Telescopes (MAST) at the Space Telescope Science Institute.  The specific observations analyzed can be accessed via \dataset[10.17909/8tdj-8n28]{https://doi.org/10.17909/8tdj-8n28}.
This survey encompasses an approximately 60 $\mathrm{arcmin}^2$ area of the sky, partially overlapping with the CANDELS GOODS-S deep field. To augment our understanding of sources in this region, we conduct a cross-match between the JADES photometric catalog \citep{2023arXiv231012340E} and the GOODS-S CANDELS catalog from \cite{2015ApJ...801...97S}. Galaxies are selected if they are detected in both surveys and have counterparts within 0.36 arcsec ($10^{-4}$ degree). To strengthen the reliability of the cross-match, we exclude galaxies with a photometric redshift difference exceeding 0.5 between the two catalogs.
We exclude sources marked with photometry flag, AGN flag and star flag from Santini's catalog. Additionally, galaxies with a non-zero {\tt FLAG} value in any band across both catalogs are removed to ensure the reliability of the images.

Physical attributes such as stellar mass ($M_*$), photometric redshift, and star formation rate (SFR) of the sources investigated in this study are taken from the GOODS-S CANDELS stellar mass catalog \citep{2015ApJ...801...97S}. 
The stellar mass and SFR are derived from the fitting of spectra energy distribution (SED), assuming an exponential declining law for star formation history (SFH): $\psi(t) \propto \exp{(-t/\tau)}$ and Chabrier initial stellar mass function \citep[IMF,][]{2003PASP..115..763C}. 
Photometric redshift comes from \cite{Dahlen_2013} with $\lvert \Delta z / (1 + z_{\mathrm{spec}}) \rvert \sim 0.01$.

In this study, we focus on star-forming galaxies, with quiescent galaxies excluded based on the criterion $\mathrm{sSFR} < 1/3t_{\mathrm{Hubble}}$ \citep{2024ApJ...961..163B}. To further ensure the purity of our sample and avoid contamination from merging systems, we impose constraints on the morphological parameter known as the asymmetry index \citep{2019MNRAS.483.4140R}. This index quantifies the degree of regularity and symmetry in the shape of galaxies. 
We apply a cutoff at $\mathrm{Asymmetry} = 0.35$ \citep{2003ApJS..147....1C, 2023ApJ...954..113Y} to identify and exclude merging systems, ensuring that our sample consists primarily of well-defined and non-merging galaxies suitable for size investigation. Finally we obtain a sample of 6384 star-forming galaxies with $M_* > 10^8 \mathrm{M}_\odot$, and 1899 of them have $M_* > 10^9 \mathrm{M}_\odot$ and redshift less than 3.5 (which are used in Section \ref{sec:3.1}).

\subsection{Method and measurement} \label{sec:2.2}


To investigate the sizes of galaxies at different bands, it is crucial to eliminate the effect of different point spread functions (PSF) at different bands.  To achieve this, we utilize the model PSF derived from \cite{ji2023jades}, which are generated with package {\sc Webbpsf} \citep[v1.1.1,][]{Perrin2012, Perrin2014}.
The \texttt{photutils} package is employed to generate a convolution kernel, which involves matching the PSF of two bands and convolving high-resolution images to produce low-resolution images, specifically targeting the long-wavelength images (F444W in the case of JADES). 
This process harmonizes the resolution of images across different bands, enabling a comparative analysis of galaxy characteristics at various rest-frame wavelengths.

To measure sizes of galaxies and obtain best fitted light curve model of them, we make use of {\sc Galfit}
\citep{2002AJ....124..266P, Peng_2010}. We fit galaxy profile with a \sersic\ profile model: 
\begin{equation}
    \Sigma (r) = \Sigma_{\mathrm{e}} \exp{[-\kappa ((\frac{r}{R_{\mathrm{e}}})^{1/n} - 1)]}, 
\end{equation}
where $\Sigma(r)$ is the light surface density, $n$ is the \sersic\ index, and $\Sigma_{\rm e}$ is the light surface density at half-light radius. $R_{\mathrm{e}}$ represents the 2D half-light radius measured along the semi-major axis. All radii utilized in this paper are the half-light radii fitted using {\sc Galfit}.
It should to be noted that uncertainties from {\sc Galfit} have been shown to be underestimated \citep{Haussler_2007}.
The segmentation map, to distinguish sources with others, and error maps are provided by JADES group \citep{2023arXiv231012340E}.
We analyze the nine bands: F090W, F115W, F150W, F200W, F277W, F335M, F356W, F410M, and F444W to determine the sizes of galaxies and exclude sources that fail to fit. Approximately $\sim 10\%$ of the sources are removed in this process. We have also examined the fitting results for some galaxies to ensure that our fitting is reliable, even for the low-surface brightness galaxies.


As mentioned in Section \ref{sec:2.1}, we only use the asymmetry index to exclude the mergers.   We therefore employ the \texttt{statmorph-csst} code \citep{2023ApJ...954..113Y} to calculate the asymmetry index based on F444W band.   Non-parametric morphology indicators can provide a powerful way to characterize the morphological properties of galaxies without presupposing any specific model of the luminosity distribution.

\section{results}\label{sec:result}
\subsection{Size variation with rest-frame wavelength} \label{sec:3.1}

\begin{figure*}[ht]
    \hspace*{\fill}
    \includegraphics[width=\textwidth]
    {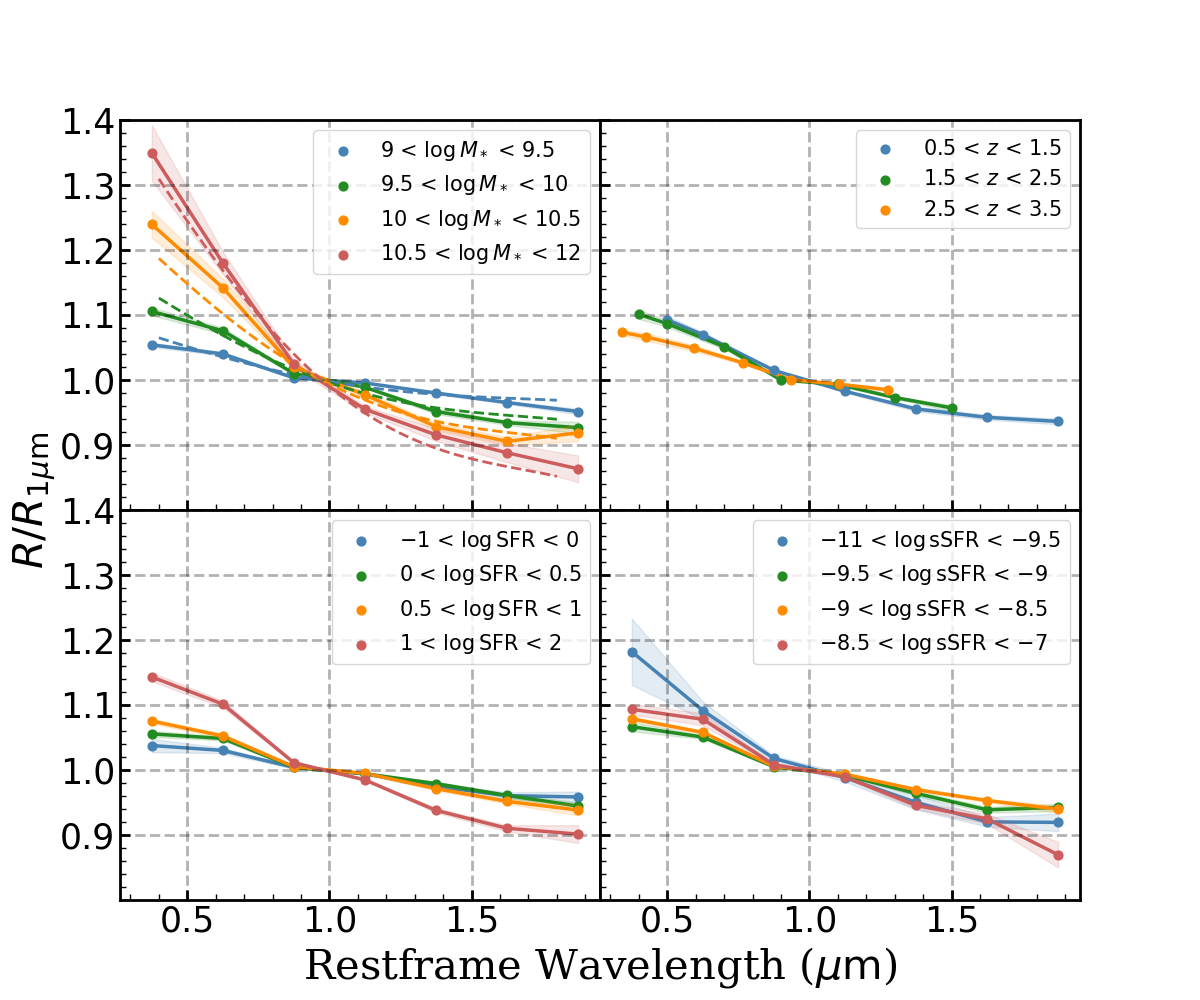}
    \hspace*{\fill}
    \caption{Size variation with rest-frame wavelength by separating galaxies according to their different physical properties, including stellar mass, redshift, SFR and specific SFR.
    In each panel, all of solid lines are median size of samples in each bins, and the shaded regions show the errors measured from the bootstrap method. 
    The top-left panel shows size variation in different stellar mass bins. Dashed lines in this panel are the best-fitted model to characterize the size variation for galaxies on both rest-frame wavelength and stellar mass, shown in Equation \ref{eq.model}. Here for all the four panels, we only use galaxies with redshift $z < 3.5$. The units of stellar mass in the figure are $\mathrm{M}_{\odot}$, and the units of SFR and sSFR are $\mathrm{M}_{\odot} \mathrm{yr}^{-1}$ and $\mathrm{yr}^{-1}$.
   }
    \label{Fig.2}
\end{figure*}

As we mentioned in the introduction, many works found that galaxies show different sizes at different wavelengths \citep{2014ApJ...788...28V, 2024MNRAS.527.6110O, ji2024jades, Suess_2022, nedkova2024uvcandelsroleduststellar, martorano2024sizemassrelationrestframe15mum}. 
By converting the observed wavelengths to rest-frame wavelengths, we obtain the sizes of each individual galaxies at a set of rest-frame wavelengths.
Through this approach, we aim to statistically analyze how size varies with rest-frame wavelength and explore correlations with other physical properties within this expanded sample. Our focus in this section is on massive galaxies with $\log M_*/\mathrm{M}_{\odot} > 9$.



We have examined that sizes of the majority of galaxies exhibit a smooth variation with wavelength, even for individual galaxies. This smooth variation allows us to estimate the size at any desired rest-frame wavelength through interpolation. 
We note that for the sizes in two adjacent bands, a linear approximation is quite effective \citep{2024ApJ...962..176W, 2021MNRAS.506..928N}.
Therefore, we employ linear interpolation to estimate sizes.
Specifically, for a given galaxy, we match the two bands that enclose the desired rest-frame wavelength and perform linear interpolation to determine its size at that wavelength. 

To standardize size variations across different wavelength bins, we calculate the sizes of all galaxies at $1 \mu \mathrm{m}$ in rest-frame, denoted as $R_{1 \mu \mathrm{m}}$. Our focus lies in examining the ratios of size to $R_{1 \mu \mathrm{m}}$ within our sample and studying their variations. 
Since direct interpolation for $1 \mu \mathrm{m}$ size is not feasible for galaxies at $z >3.5$, we only use galaxies with redshift $z < 3.5$ here.
To exclude galaxies with catastrophic measurements of size, in this subsection we exclude galaxies with the uncertainty of $R_{1 \mu \mathrm{m}}$ to be larger than 60\%
and make a $4\sigma$ clip iteratively to exclude the outliers in the measurements of $R_{1 \mu \mathrm{m}}$, which in total remove about $10\%$ of sources.  

\subsubsection{Dependence on galaxy properties} \label{sec:3.1.1}

Figure \ref{Fig.2} shows the size variation with rest-frame wavelength for galaxies of different properties. 
We normalize the sizes of all galaxies across each band using $R_{1 \mu \mathrm{m}}$. Based on their physical properties, we categorize our sample into different bins and show the median size ratio ($R/R_{1 \mu \mathrm{m}}$) as a function of rest-frame wavelength in Figure \ref{Fig.2}.  Errors of each bin denoted with the shaded region are calculated with bootstrap method with 100 times random samplings with replacement. Throughout this work, the errors in such kinds of plots are all measured in this way.  

As illustrated in Figure \ref{Fig.2}, galaxies at shorter wavelengths show larger size across the full range of wavelengths (from UV-to-NIR) we considered here. 
It is qualitatively consistent with \cite{2014MNRAS.441.1340V}.
In addition, our analysis reveals that stellar mass emerges as the most influential factor affecting size variations with wavelength, as observed in the top-left panel. For low-mass galaxies, such as those with $M_* < 10^{9.5}\mathrm{M}_{\odot}$, the median size 
variation 
between $0.5 \ \mu \mathrm{m}$ and $2 \ \mu \mathrm{m}$ bands remains within approximately $10\%$. In contrast, for massive galaxies ($M_* > 10^{10.5} \mathrm{M}_{\odot}$), 
the size of $0.5 \ \mu \mathrm{m}$ is overall $\sim40\%$ larger than that of $2 \ \mu \mathrm{m}$.
Notably, among the most massive galaxies ($M_* > 10^{10.5} \mathrm{M}_{\odot}$), $20\%$ of the sample have doubled their sizes at $0.5 \ \mu \mathrm{m}$ with respect to sizes at $2 \ \mu \mathrm{m}$. These findings underscore the critical role of stellar mass in determining sizes across different wavelengths. Furthermore, the top-left panel also illustrates the trend of size variations. Sizes exhibit rapid changes at shorter wavelengths, particularly noticeable for massive galaxies where sizes can vary by over $30\%$ between approximately $0.4 \ \mu \mathrm{m}$ and $1 \ \mu \mathrm{m}$. In contrast, the difference between $R_{1 \mu \mathrm{m}}$ and $R_{2 \mu \mathrm{m}}$ is only around $10\%$, highlighting a less pronounced variation in longer wavelength bands. 

To assess whether dust attenuation significantly affects our results, we have also examined the effect of the axis ratio on our results, because dust attenuation is expected to correlate with inclination \citep{Wang_2018_irx,lu2022chocolatechipcookiemodel}.
We found that our conclusions remain consistent when separating galaxies with lower and higher axis ratios, suggesting that dust may not be a dominant factor. However, some studies found that at $z \sim 1-2$, dust extinction begins to show weak or even no dependence on galaxy minor-to-major axis ratio \citep{Lorenz_2023, Zhang_2023}. It appears that there is still no consensus on this matter.
Objects in \cite{2015ApJ...801...97S} catalog with mid/far-IR measurement might be a great testbed to investigate the effect of dust attenuation, but this sample is too small for statistical analysis. We note that the measurement of SFR might also be affected by effects coupled with dust attenuation.


In contrast to the significant impact of stellar mass, Figure \ref{Fig.2} indicates that redshift and star formation activity have less critical roles in size variation, as depicted in the other panels. The top-right panel demonstrates that galaxies with $z > 2.5$ exhibit slight variations, particularly at shorter wavelengths, suggesting a very gentle influence of redshift on size variation. However, this effect is not substantial enough to decisively attribute size changes to redshift alone, 
the different mass distribution in each redshift bin might also contribute to this outcome.

The bottom-left panel reveals size variations across different SFR bins. Galaxies with $\mathrm{SFR} > 1$ $\mathrm{M}_{\odot} {\rm yr}^{-1}$ show sizes in $0.5 \ \mu \mathrm{m}$ that are approximately $25\%$ larger than those in $2 \ \mu \mathrm{m}$. Conversely, galaxies with $\mathrm{SFR} < 1$ $\mathrm{M}_{\odot} {\rm yr}^{-1}$ exhibit gentler size variations. Since there is a strong correlation between SFR and stellar mass, we examine the specific SFR ($\mathrm{sSFR} = \mathrm{SFR} / M_*$) in the bottom-right panel, to disentangle the effects of stellar mass. Controlling for stellar mass reveals that star formation activity seems to have little impact on size variation. This finding suggests that once the influence of stellar mass is accounted for, the correlation between star formation activity and size variation diminishes.

We conclude that stellar mass is the most important parameter we have investigated to determine size variation with rest-frame wavelength, rather than redshift and sSFR.  Motivated by this, we provide an analytical formula to describe the size variation with wavelength, which can be treated as a rough correction when the sizes of the needed bands are not available.  We fit the mass dependence of size variation in the top-left panel of Figure \ref{Fig.2} with a polynomial function, which can be written as
\begin{equation}
\begin{split}
    \frac{R_{\mathrm{eff}}(\lambda)}{R_{1 \mu \mathrm{m}}} = 1 + \log \frac{ M_*}{10^{8.72}\mathrm{M}_{\odot}} \cdot (-0.044 x^4 + \\
    0.173 x^3 - 0.134 x^2 - 0.223 x + 0.223),
\end{split}
\end{equation}
where $x$ is $\lambda / 1 \mu \mathrm{m}$. This formula is shown in dashed lines in the top-left panel of Figure \ref{Fig.2}, which overall captures the size dependence on both stellar mass and wavelength roughly within 1$\sigma$ region. 

Since size variation with wavelength reflects the different distributions for stellar populations of different ages if ignoring dust attenuation, the decrease of size with increasing wavelength might indicate that SF galaxies are overall growing their sizes.  This also means most of star-forming galaxies have a negative radial gradient of stellar age.

\subsubsection{Environmental effect} \label{sec:3.1.2}

\begin{figure*}[ht]
    \centering
    \includegraphics[scale=0.45]{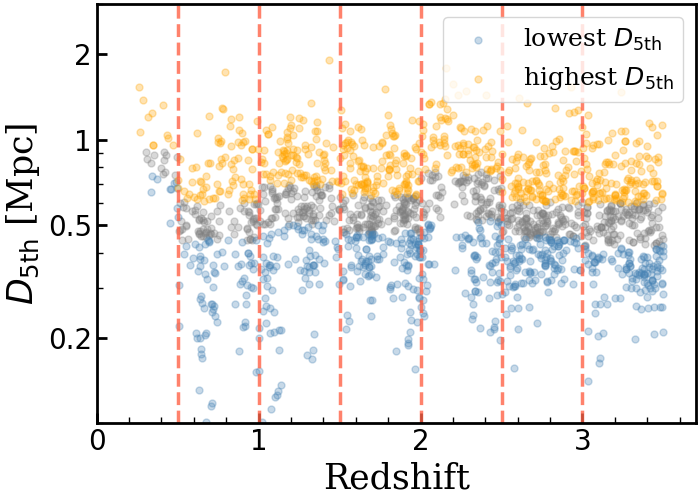}
    \includegraphics[scale=0.46]{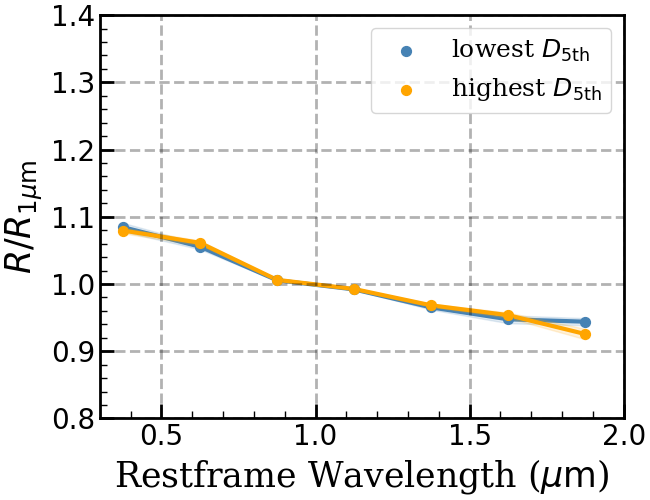}
    \caption{The dependence of size variation on galaxy environment. 
    The left panel shows the variation of $D_{\mathrm{5th}}$ with redshift, the blue and orange dots represent the lowest and highest one-third of $D_{\mathrm{5th}}$ galaxies in each redshift bin, respectively.
    In the right panel, the impact of the environment on size variation is depicted. The blue line represents galaxies with lowest $D_{\mathrm{5th}}$ in the left panel, indicating a denser environment. The orange line represents galaxies with highest $D_{\mathrm{5th}}$. 
    }
    \label{Fig.insert}
\end{figure*}

\begin{figure*}[ht]
    \hspace*{\fill}
    \includegraphics[width=\textwidth]{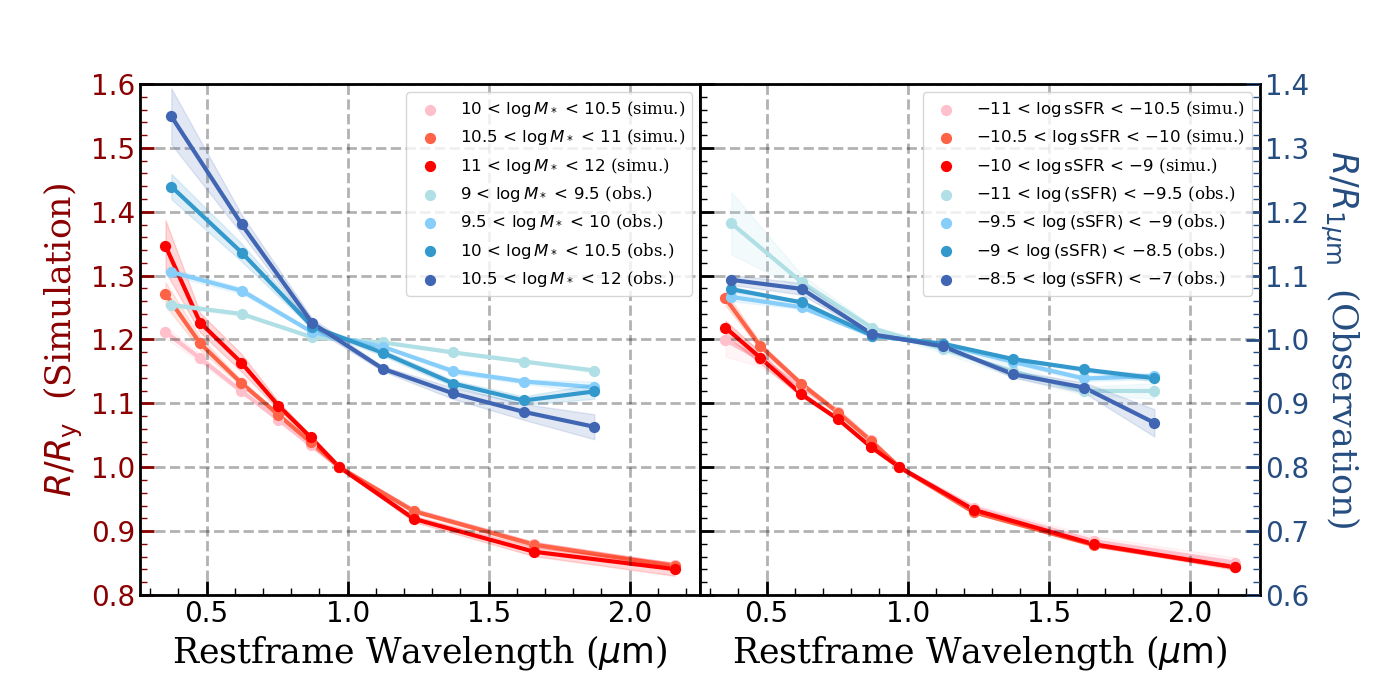}
    \hspace*{\fill}
    \caption{ Comparison of size variation with wavelength between the observation and the TNG50 simulation.  
    In each panel, the blue series of lines represent our observational results ($z < 3.5$), corresponding to the axis and label on the right. The red series of lines depict the size variation of TNG50 simulated data at redshift of zero \citep{2024arXiv240104225B}, corresponding to the axis and label on the left. The left panel illustrates the dependence of size variation on stellar mass, while the right panel shows the same but for sSFR. The units of stellar mass in the figure are $\mathrm{M}_{\odot}$, and the units of SFR and sSFR are $\mathrm{M}_{\odot} \mathrm{yr}^{-1}$ and $\mathrm{yr}^{-1}$.
    }
    \label{Fig.3}
\end{figure*}


Considering the influence of environmental effects on size variation may be crucial,  since the mass assembly of star-forming galaxies is mostly driven by the smooth accretion of gas from their local environment \citep[e.g.][]{1994MNRAS.271..676L, 2002ApJ...571....1M, 2008A&ARv..15..189S, 2009AAS...21331306K, 2010ApJ...718.1001B, 2012A&A...544A..68L, 2013MNRAS.430.1051C, 2014A&ARv..22...71S, 2015MNRAS.449...49R}.   
\cite{2014ApJ...788...28V} highlighted an intriguing relationship between the sizes of star-forming galaxies and the virial radius of their host dark matter halo, which is likely originated from the conservation of angular momentum \citep{1980MNRAS.193..189F, 1998MNRAS.295..319M, 2009MNRAS.396..121D, 2018MNRAS.473.2714S}. This suggests that the local environment may play a role in shaping the sizes or size growth of star-forming galaxies. 


Studying the galactic environment, especially for high-$z$ galaxies, poses several challenges. Firstly, detecting high-$z$ galaxies can be difficult, potentially leading to an underestimation of their number density. Secondly, establishing a consistent definition to characterize the environment of both low-$z$ and high-$z$ galaxies is a persistent issue. To address these challenges, we utilize the concept of $D_{\mathrm{5th}}$ \citep{Etherington_2015}, which quantifies the number density of galaxies surrounding a specific target. This measure represents the projected distance to the 5th nearest neighbor within a small redshift interval, denoted as $\mathrm{d}z$.  For simplicity, we only consider the galaxies (as well as neighbour galaxies) with stellar mass greater than $10^9\mathrm{M}_{\odot}$ and redshift less than 3.5. 
\cite{Etherington_2015} have tested the feasibility to use the photometric redshift to quantify the galaxy environment with respect to the spectroscopic redshift. They found good correlation between the photometric environments and benchmark environments with Spearman rank correlation coefficient of 0.4, adopting the optimal parameter values at a redshift uncertainty of 0.1.  
In this work, we set $\mathrm{d}z=0.05$, which is slightly larger than the uncertainty of photo-$z$ \citep{Dahlen_2013}. However, we have tested that a smaller or a higher $\mathrm{d}z$ does not alter the conclusion in the subsection. 

To mitigate the complexities associated with redshift dependence, we divide our sample into three subsamples with equal number of galaxies according to $D_{\mathrm{5th}}$ at given small redshift bins in Figure \ref{Fig.insert}. 
This approach enables us to statistically analyze the influence of the galactic environment by eliminating the redshift-dependent effects. 
We focus on comparing the size variation of galaxies with the lowest and highest $D_{\mathrm{5th}}$. 
Additionally, we also control the stellar mass distribution of these two subsamples, to eliminate the effect of stellar mass. 

The result is shown in the right panel of Figure \ref{Fig.insert}. As can be seen, there is almost no differences in the size variations between these two bins after controlling the stellar mass. This result indicates that the environment does not appear to play a role in determining the size variation with wavelength.  

\subsubsection{Comparison to TNG simulation}  \label{sec:3.1.3}

The stellar mass dependence of size variation observed in Section \ref{sec:3.1.1} is also evident in the TNG50-SKIRT Atlas simulation \citep{baes2024tng50skirt}. In this subsection, we perform a comparison between the observational and TNG50-SKIRT results.  

The TSA, short for the TNG50 Stellar-mass Atlas, is a comprehensive collection of synthetic images comprising a carefully selected sample of 1154 galaxies extracted from the TNG50 simulation \citep{2019MNRAS.490.3196P, 2019MNRAS.490.3234N}. As the highest-resolution version of the IllustrisTNG simulation, TNG50 has a baryonic mass resolution of $8.5 \times 10^4 \mathrm{M}_{\odot}$ and a spatial resolution of 70 to 140 pc \citep{2018MNRAS.480.5113M, 2018MNRAS.477.1206N, 2018MNRAS.475..624N, 2018MNRAS.475..648P, Springel_2017}. 
The TSA generates noise-free images for each galaxy from five different random observer positions, omitting the convolution with the PSF.
By comparing the ``observed'' images of different bands generated with the SKIRT radiative transfer code by \cite{2015A&C.....9...20C}, we can perform a relative fair comparison with our observational results.

Figure \ref{Fig.3} shows the comparison between our result and TNG50 simulation shown in \cite{2024arXiv240104225B}.  
The simulated data shown in this figure are directly taken from \cite{2024arXiv240104225B}, encompassing the u, g, r, i, z, y, J, H, and $\mathrm{K_s}$ bands. This wavelength range is similar to that of our galaxies at rest-frame in JADES. We note that this set of galaxies has stellar masses between $10^{9.8} \mathrm{M_{\odot}}$ and $10^{12} \mathrm{M_{\odot}}$, which is generally more massive than the galaxies of our sample.
In this figure, sizes of galaxies in simulation are normalized to $y$-band, which is approximately $1 \ \mu \mathrm{m}$. 
Even though these results are specific to the $z = 0$ slice of TNG50, we can still observe some similar trends, since redshift does not play an important role (Figure \ref{Fig.2}). The left panel illustrates the variation in the size ratio $R/R_{\rm y}$ when the sample is divided into different stellar mass intervals. Because of the smaller stellar mass range for TNG50 sample with respect to JADES sample, the stellar mass dependence is not as distinct as seen in our observations. However, there is still a noticeable trend of size varying more rapidly at higher masses. In addition, the right panel of Figure \ref{Fig.3} indicates little or no correlation between size variation and sSFR, which is entirely consistent with the findings in Section \ref{sec:3.1.1}.

It is essential to note that dust significantly influences size variations, as demonstrated in simulations \citep{2024arXiv240104225B}. The presence of dust can impact the observed sizes due to the strong dependence of attenuation on wavelength.  
\cite{nedkova2024uvcandelsroleduststellar} found that dust attenuation can cause size differences at various rest-frame wavelengths \citep[also investigated in][]{2022MNRAS.511.5475M}.  However, as demonstrated in \cite{2024arXiv240104225B}, if the dust content is significant enough, an increase in dust tends to reduce the size differences across different wavelengths, suggesting that the effects of dust may not be monotonic.
If this is true, considering that massive galaxies often harbor higher dust percentages, they might exhibit a steeper slope of size variation intrinsically compared to what is depicted in Figure \ref{Fig.2}. 
There is currently no consensus on the extent to which dust affects size variation with wavelength. Our forthcoming work will focus on quantifying the effects of dust on size variation through SED fitting.
This approach will provide a more precise understanding of how dust influences size variations in the observations. 
In this work, we do not yet consider the effects of dust. As shown by \cite{2024arXiv240104225B}, stellar population gradients are the primary factor ($\sim 80\%$) driving the wavelength dependence of the effective radius. 
Later in Section \ref{sec:discuss}, we will show that the typical size variation with wavelength provides important information about size evolution of galaxies on different timescales.

\subsection{The evolution of mass-size relation} \label{sec:3.2}

\begin{figure*}[ht]
    \hspace*{\fill}
    \includegraphics[width=\textwidth]{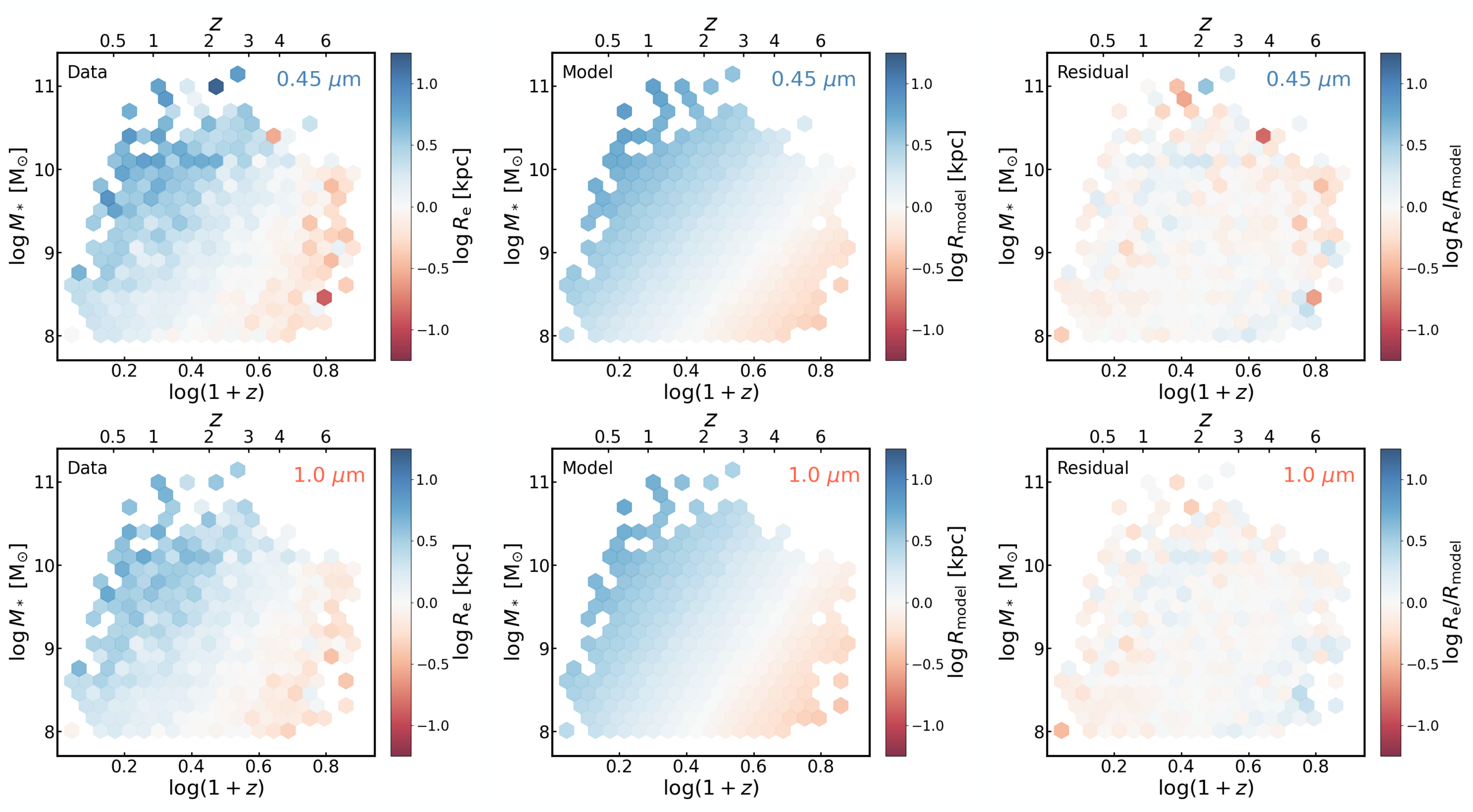}
    \hspace*{\fill}
    \caption{ The mapping of galaxy size at rest-frame $0.45 \mu \mathrm{m}$ and  $1.0 \mu \mathrm{m}$ on the stellar mass vs. redshift diagram. 
    For the left column, color represents the median effective radius of galaxies within hexagon bins.
    The middle column shows our best-fitted models, and the right column show the residual maps.}
    \label{Fig.4}
\end{figure*}

\begin{figure*}[ht]
    \hspace*{\fill}
    \includegraphics[width=\textwidth]{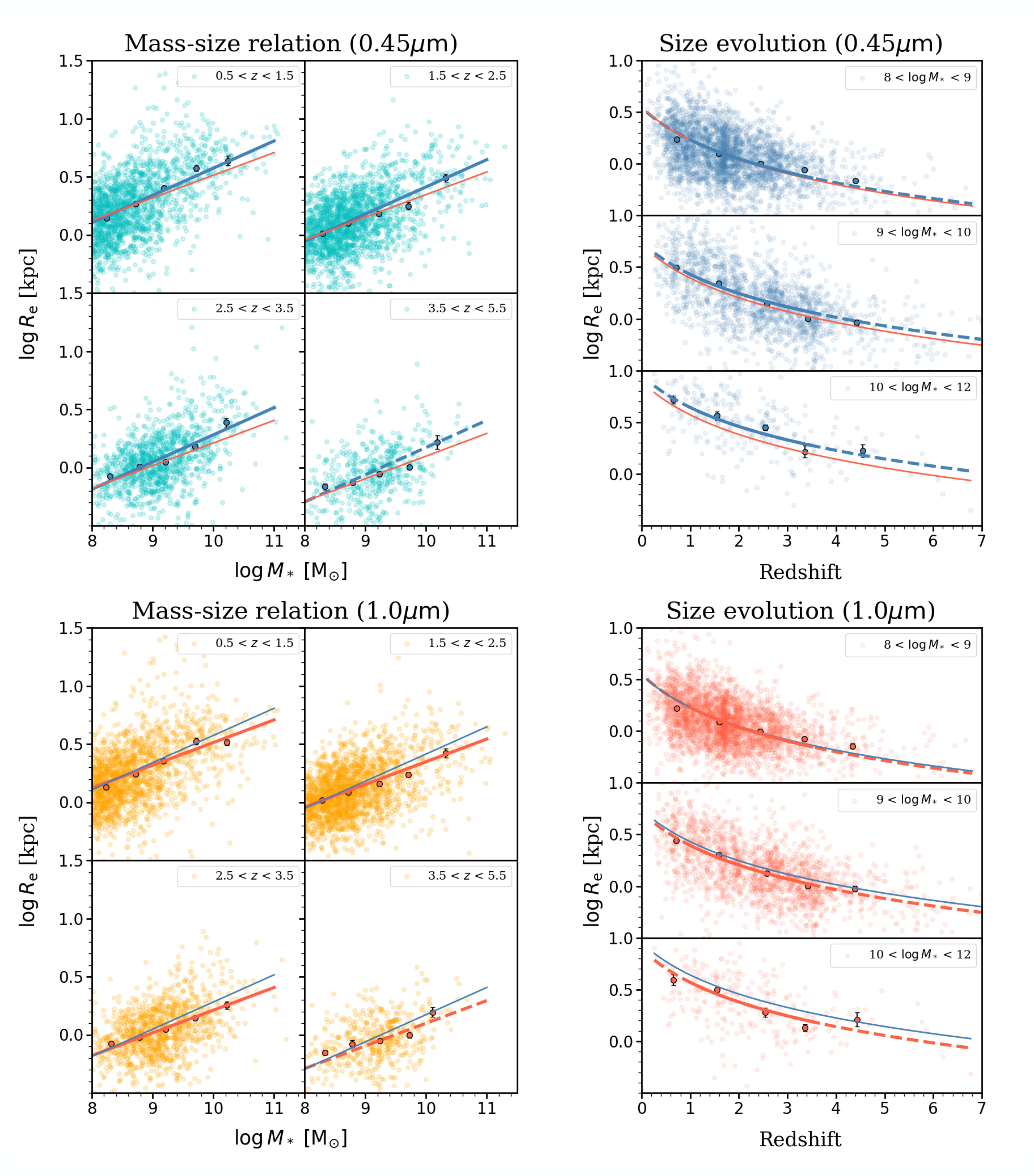}
    \hspace*{\fill}
    \caption{The evolution of mass-size relation and size evolution for galaxies of different mass intervals at wavelength of $0.45 \ \mu \mathrm{m}$ (top group of plots) and $1.0 \ \mu \mathrm{m}$ (bottom group of plots).
    In each group, the left column shows mass-size relation at four different redshift bins, and the right column shows size evolution for galaxies of different stellar mass intervals. 
    In each panel, we show the best-fit model with Equation \ref{eq.model} in solid and/or dashed lines.  The dashed lines or dashed parts highlight the redshift intervals that are not used in the fittings, and show the interpolations of best-fit model.
    In each panel, dots with black edge show the median size in each mass/redshift bins, which are approximately consistent with the fitted lines. The units of stellar mass in the figure are $\mathrm{M}_{\odot}$.
    }
    \label{Fig.5}
\end{figure*}

\begin{figure*}[ht]
    \hspace*{\fill}
    \includegraphics[width=\textwidth]{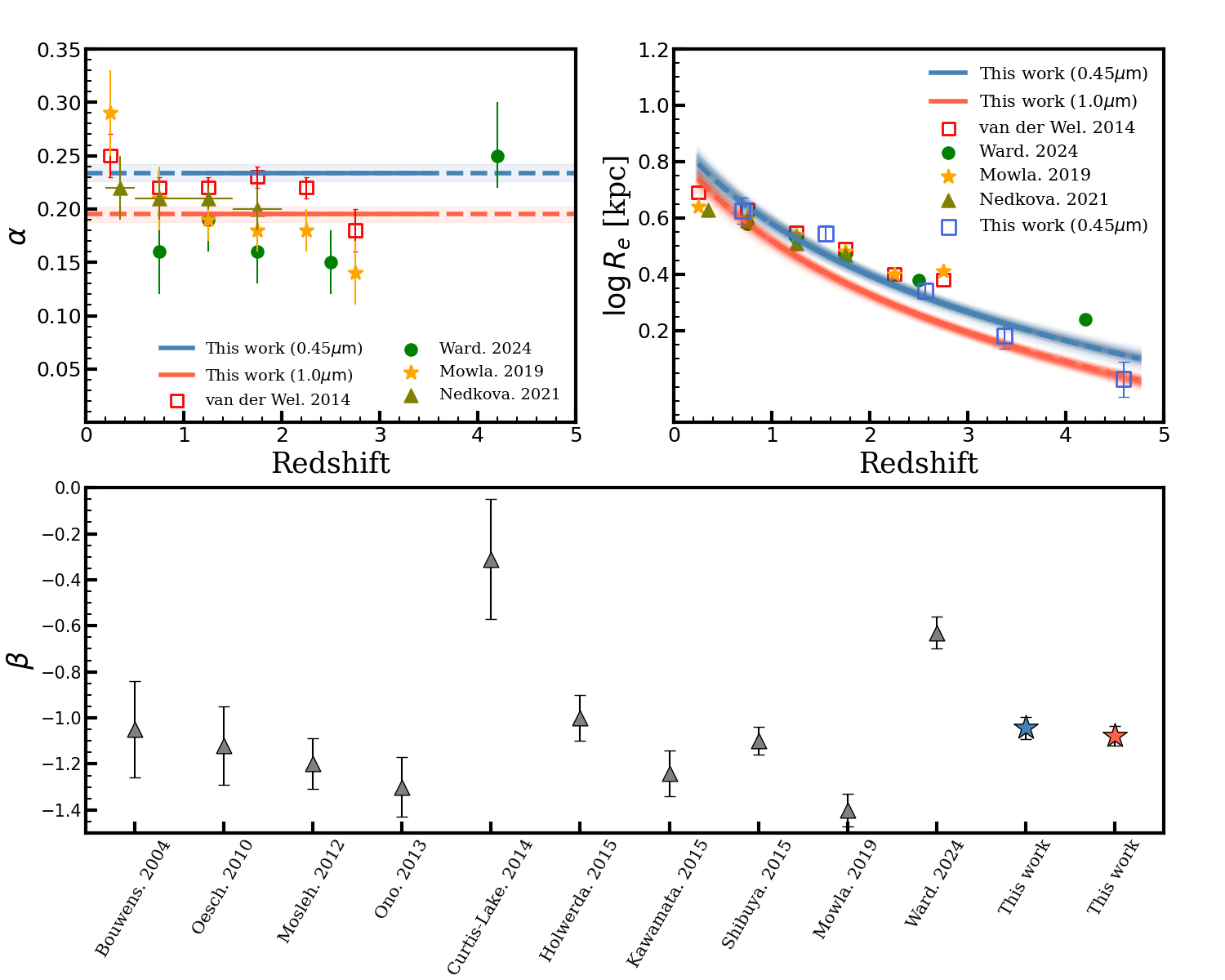}
    \hspace*{\fill}
    \caption{Comparison of the fitting result with previous works. The top-left panel compares the slope $\alpha$ of the mass-size relation $R_{\mathrm{e}} \propto M_*^{\alpha}$ to the $\alpha$ measured from the literature as a function of redshift. 
    The blue line shows $\alpha$ for $0.45 \ \mu \mathrm{m}$, and the red line shows $\alpha$ for $1.0 \ \mu \mathrm{m}$. In the top-right panel, the size evolution with redshift is depicted at stellar mass of $10^{10} \mathrm{M}_{\odot}$. Here, the blue and red lines denote the best-fit models at $0.45 \ \mu \mathrm{m}$ and $1.0 \ \mu \mathrm{m}$, respectively. The blue squares represent the median size of JADES galaxies with the stellar mass between 9.8 and 10.5 (with median value of $10^{10.01}{\rm M}_{\odot}$) for $0.45 \ \mu \mathrm{m}$ at different redshift bins. Other markers are extracted from the previous works, showing the evolution of optical size for galaxies of the same stellar mass.The bottom panel compares our $\beta$ of $R_{\mathrm{e}} \propto (1+z)^{\beta}$ with previous works.}
    \label{Fig.6}
\end{figure*}


The evolution of galaxy sizes and the mass-size relation have been studied by many authors \citep{2014ApJ...788...28V, 2015ApJ...813...23V, 2019ApJ...872L..13M, 2019ApJ...880...57M, Miller_2019}. Extending their findings to longer wavelength bands is crucial as different rest-frame wavelengths offer distinct insights into the stellar populations of galaxies.  
As discussed in Section \ref{sec:3.1.1}, strong size variation with rest-frame wavelength suggests that the studies of mass-size relation evolution for different wavelengths are required.  
Some studies at longer wavelengths have provided insights into this issue \citep{martorano2024sizemassrelationrestframe15mum, Suess_2022}. However, researches on this topic remain limited.

We then choose two rest-frame wavelengths, $0.45 \ \mu \mathrm{m}$ and $1 \ \mu \mathrm{m}$, as a representative of the relative distribution of young and old stellar populations for galaxies. 
For individual galaxies, the sizes at these two given wavelengths are obtained by interpolation (see Section \ref{sec:3.1}).  
We take the size of shortest band of JADES as the size of $0.45 \ \mu \mathrm{m}$ if all the bands are greater than  $0.45 \ \mu \mathrm{m}$ in  rest-frame. In the same way, we take the size of longest band of JADES as the size of $1.0 \ \mu \mathrm{m}$ in case that all the bands are shorter than  $1.0 \ \mu \mathrm{m}$ in rest-frame. 
In this section, we will investigate how size evolves with mass and redshift at these two wavelengths.


The left column of panels in Figure \ref{Fig.4} shows how the size of galaxies varies with redshift and stellar mass at these two rest-frame wavelengths. The color-coding of each hexagonal bin shows the median size of the corresponding region on the stellar mass and redshift diagram.
These panels illustrate the overall trend of size evolution with redshift and stellar mass  of galaxies with size uncertainty less than $60\%$ at two different rest-frame wavelengths. Both bands reveal a similar pattern: size increases with mass but decreases with rising redshift. However, a distinction exists between these two wavelengths. Size variation appears to be steeper at $0.45 \ \mu \mathrm{m}$ compared to $1.0 \ \mu \mathrm{m}$. Notably, we observe larger sizes at shorter wavelengths for high-mass galaxies than low-mass ones, which aligns with the trends depicted earlier in Figure \ref{Fig.2}.


Previous studies have shown that the slope of the mass-size relation remains nearly constant with redshift, with the intercept varying only \citep{2014ApJ...788...28V, 2019ApJ...880...57M, 2024ApJ...962..176W}, although this relation may exhibit non-linearity at extremely high masses (around $M_* > 10^{11} \mathrm{M}_{\odot}$) \citep{2019ApJ...872L..13M}. Consequently, if the mass-size relation of star-forming galaxies can be described as linear with a constant slope and a redshift-dependent intercept, it implies that size evolution is also mass-independent, and exponent $\beta$ in $R_{\mathrm{e}} \propto (1+z)^{\beta}$ remains constant across different masses. 
These trends are evident in our results, as shown in Figure \ref{Fig.4}. Notably, the effective radius, mass, and redshift align closely in logarithmic space, indicating that the slopes associated with these parameters are indeed independent of each other. 
Motivated by this, we model the size as a function of both stellar mass and redshift with a simple linear function in logarithmic space: 
\begin{equation}\label{eq.model}
    \log\frac{R_{\rm e}}{\rm kpc} = \alpha \log\frac{M_*}{\rm \mathrm{M}_{\odot}} + \beta \log{(1+z)} + k,
\end{equation}
where $\alpha$ reflects the dependence on $M_*$, $\beta$ reflects the evolution with redshift, and $k$ is the intercept. 


We employ a 2D-fitting approach with the Markov Chain Monte Carlo (MCMC) method with {\tt emcee} \citep{Foreman-Mackey-13}.
In calculating the likelihood function, we input the errors of the median sizes in each hexagonal bin indicated by the $\sigma(R) / \sqrt{N}$, where the $\sigma(R)$ is the scatter of galaxy sizes in that hexagonal bin, and $N$ is the number of galaxies in that bin.  This error includes two things, i.e. the intrinsic dispersion in size within each hexagonal bin and the true uncertainty of size measurement in the {\sc Galfit} fitting. 
To prevent a large number of low-mass galaxies from dominating the fitting process, we utilize hexagon-binned data in Figure \ref{Fig.4} with each hexagon being equally weighted in the fitting.
However, a potential issue with fitting arises due to observational limitations in wavelength and redshift. Specifically, size measurements at specific rest-frame wavelengths may not be accurate at extremely low or high redshifts because all the bands in JADES are less than $1.0 \ \mu \mathrm{m}$ or greater than $0.45 \ \mu \mathrm{m}$ at rest-frame. To address this concern, we restrict our fitting analysis to galaxies with redshifts ranging from 1 to 3.5.
For wavelength of $0.45 \ \mu \mathrm{m}$, we obtain the best-fitted parameters in Equation \ref{eq.model} as: $\alpha = 0.234 \pm 0.008$, $\beta = -1.044 \pm 0.049$, $k = -1.446 \pm 0.069$, and for wavelength of $1.0 \ \mu \mathrm{m}$, we obtain $\alpha = 0.195 \pm 0.007$, $\beta = -1.077 \pm 0.042$, $k = -1.107 \pm 0.059$. 
Additionally, to determine whether the uncertainties in stellar mass affect our results, we conducted the following test. By adding a Gaussian error with $ \sigma \sim 0.2 \ \mathrm{dex}$ to the measurements of stellar mass manually, we find that the slope $\alpha$ (of the mass-size relation) changes from 0.23 to approximately 0.18. This means that the uncertainty in mass could flatten the mass-size relation, and the real $\alpha$ should be slightly higher than what we measured here.

The middle and right columns of Figure \ref{Fig.4} present the fitted size map on the stellar mass and redshift diagram  and the corresponding residual map, respectively.  As can be seen, this simple model can very well capture the features of size variation on the diagram, leading to reasonable small residuals.
In Figure \ref{Fig.5}, we examine our 2D fits by comparing the model with the observed mass-size relation at different redshift bins and the size evolution for given mass bins, for the two wavelengths.   As can be seen, the simple model overall describes the observed mass-size relation and size evolution very well. This confirms that it is reasonable to describe size evolution with mass and redshift as a power law, and slopes of them are independent with each other \citep{2014ApJ...788...28V, 2019ApJ...880...57M, 2024ApJ...962..176W}.  It is interesting to note that galaxies out of the fitting redshift range also appear to  broadly follow the fitted relation.

The best-fit value of $\alpha$ is smaller at wavelength of $1.0 \ \mu \mathrm{m}$ than at wavelength of $0.45 \ \mu \mathrm{m}$.  This is consistent with the results of Section \ref{sec:3.1.1} that more massive galaxies have larger size ratio $R_{0.45 \mu \mathrm{m}}/R_{1.0 \mu \mathrm{m}}$.  This also indicates that more massive galaxies will increase in size more significantly compared to less massive galaxies as a given fraction of stellar mass increases, as discussed in Section \ref{sec:3.3.1}.

In Figure \ref{Fig.6}, we compare the fitted values of $\alpha$ and $\beta$ with studies from the literature. As shown in the top-left panel, the slope $\alpha$ of  mass-size relation at the two specific wavelengths broadly show good consistency with previous studies. Most studies report the values of $\alpha$ ranging from 0.15 to 0.25, with some exceeding this range, while our result is closed to \cite{2014ApJ...788...28V}.
In addition, these previous studies show that $\alpha$ nearly does not evolve with redshift (for $z<3$), which again enables our 2D model of size evolution with Equation \ref{eq.model}.  

In the top-right panel of Figure \ref{Fig.6}, we compare the size evolution at a specific intercept ($10^{10} \mathrm{M}_{\odot}$) with previous works, particularly those using HST data (\cite{2024ApJ...962..176W} used JWST data). As we can see, even with different datasets and different methods, our simple model is broadly in good agreements with previous results for the optical size.   In the bottom panel, we delve into the comparison of the slope of size evolution, i.e. $\beta$. Although there is a notable diversity among studies in the values of $\beta$, the $\beta$ obtained in this work are within the range of previous measurements \citep{2004ApJ...611L...1B, 2010ApJ...709L..21O, 2012ApJ...756L..12M, 2013ApJ...777..155O, Curtis_Lake_2016, Holwerda_2015, Kawamata_2015, 2015ApJS..219...15S, 2014ApJ...788...28V, 2019ApJ...880...57M, 2021MNRAS.506..928N, 2024ApJ...962..176W}. 
Overall, we conclude that our simple model provides a reasonable fitting to the data and exhibits good consistency with previous results.   


\subsection{Size ratio as tracers of galaxy growth}   \label{sec:3.3}

\subsubsection{Size ratio} \label{sec:3.3.1}

\begin{figure}[ht]
    \centering
    \includegraphics[scale=0.35]{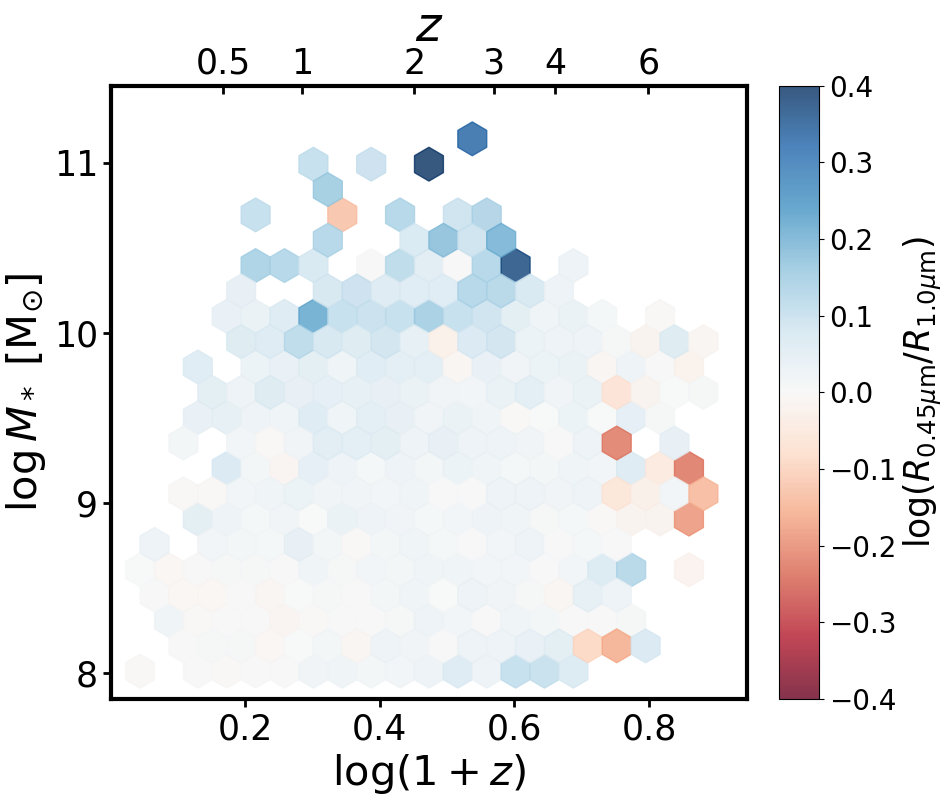}
    \caption{Size ratio between $R_{0.45 \mu \mathrm{m}}$ and $R_{1.0 \mu \mathrm{m}}$ in redshift-mass diagram.}
    \label{Fig.ratio}
\end{figure}

As previously noted, the size ratio $R_{0.45\mu \mathrm{m}} / R_{1.0\mu \mathrm{m}}$ can reflect the degree of size growth when the stellar mass of a galaxy increases by a certain proportion. We note that the size of rest-frame near-UV band shows consistent trend on the relation of size variation with wavelength. Since the UV band can be a good tracer of star formation at timescale of 100-300 Myr \citep{calzetti2012starformationrateindicators},  the size ratio $R_{0.45 \mu \mathrm{m}}/R_{1.0 \mu \mathrm{m}}$ indicates the size growth at a timescale of a few hundred Myr.
Figure \ref{Fig.ratio} shows the size ratio $R_{0.45\mu \mathrm{m}} / R_{1.0\mu \mathrm{m}}$ as a function of both stellar mass and redshift.  Massive galaxies exhibit higher size ratios, consistent with the observations illustrated in Figure \ref{Fig.2}.  This suggests that massive galaxies tend to have more extended newly formed stars with respect to their stellar disk, than less massive galaxies.  In other words, massive galaxies grow faster in size with increasing a given fraction of stellar mass than less massive galaxies. 

We note that low-mass galaxies tend to show a size ratio of approximately 1 as shown in Figure \ref{Fig.ratio}. This suggests that the ``inside-out'' growth mode in low-mass galaxies become ambiguous, which is consistent with previous findings at low-redshift that the radial profile of sSFR gradually flattens towards low stellar mass end \citep{Wang-19, 2018MNRAS.480.2544R, 2023ApJ...945..117A}.



Dust tends to attenuate the light of shorter wavelengths \citep[see][]{2020ARA&A..58..529S}. 
We can not fully decouple the role of dust in the present work.   
As mentioned above, the influence of dust on the size variation of galaxies with wavelength is quite complex.
In Section \ref{sec:discuss}, we validate this interpretation with an independent approach, illustrating that the size ratio effectively tracks the evolutionary trajectory of individual galaxies on the mass-size plane, i.e. the size growth.  


It is interesting to point out that some galaxies with low-mass and high-z have size ratio $R_{0.45\mu \mathrm{m}} / R_{1.0\mu \mathrm{m}}$ less than 1 at $z > 4$. This implies that the star formation in these galaxies may be centrally concentrated, leading to a negative growth in size, 
which also indicate these high redshift galaxies might have a positive stellar age gradient.
Consistent with this, by investigating the spatially resolved emission-line properties of 63 low-mass galaxies at $4<z<10$ from JADES, \cite{tripodi2024spatially} found strong negative gradients for the equivalent width of H$\beta$ at the significance of 7$\sigma$. In addition, being equivalent width, it is in principle not sensitive to the dust attenuation. 
This trend persists even after removing AGN candidates, so they proposed that the sample is dominated by active central star formation, along with inverted metallicity gradients attributed to the recent accretion of pristine gas.  
Using the images of different wavelengths, we here show quite similar results for the low-mass and high-$z$ galaxies in JADES.    
These galaxies are rapidly assembling their central stellar cores (or bulges), which are  likely suffering from wet ``compaction'' process, as proposed by many works \citep[e.g.][]{2013ApJ...776...63F, 2015MNRAS.450.2327Z, 2016ApJ...827L..32B, Tacchella_2016}. 



\subsubsection{Typical size growth trajectory for star-forming galaxies} \label{sec:discuss}
\begin{figure*}[ht]
    \hspace*{\fill}
    \includegraphics[width=\textwidth]{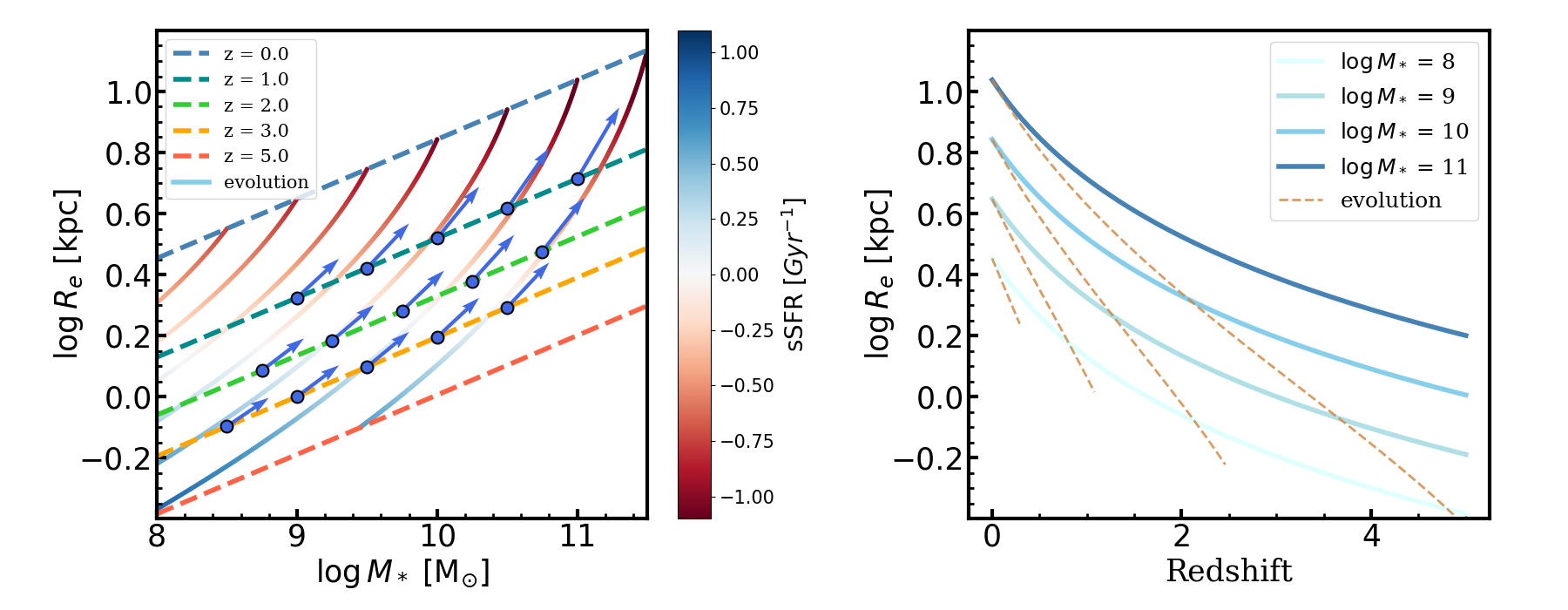}
    \hspace*{\fill}
    \caption{The typical size evolution of individual galaxies of different stellar masses. 
    The left panel shows the typical evolution locus of individual galaxies on the mass-size diagram. 
    Dashed lines are mass-size relations of different redshifts for wavelength of $1.0 \mu \mathrm{m}$, as modelled with Equation \ref{eq.model}. 
    Solid lines show typical growth track of individual galaxies. Blue dots and arrows indicate the slope of growth track at these points on mass-size map. 
    In the right panel, the blue lines show size evolution of galaxy population with different stellar masses, while dashed lines are typical size growth track with redshift for individual galaxies. The units of stellar mass in the figure are $\mathrm{M}_{\odot}$.}
    \label{Fig.8}
\end{figure*}

\begin{figure}[ht]
    \centering
    \includegraphics[scale=0.4]{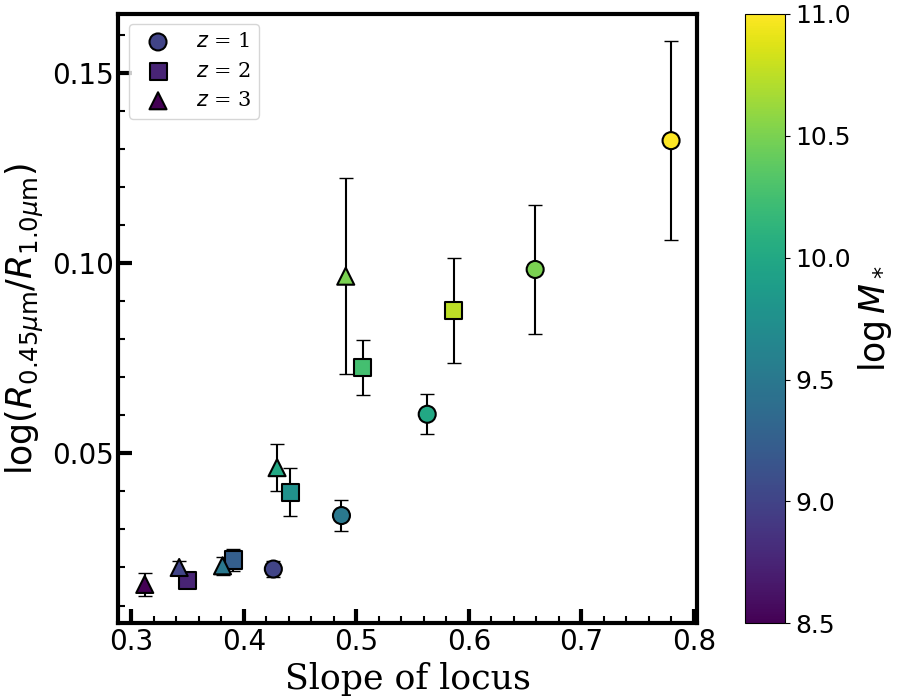}
    \caption{Comparison between slope of growth track for individual galaxies (shown in blue dots in the left panel of Figure \ref{Fig.8}) and the median size ratio of JADES galaxies that are at similar stellar mass and redshift.  Color shows the median stellar mass of them.}
    \label{Fig.ratio_slope}
\end{figure}


Combining the mass-size relation with the evolution of star formation main sequence (SFMS) offers numerous advantages, with one of the most significant being the ability to discern the typical  growth trajectory of individual galaxy sizes over time. 
For a star-forming galaxy of a given stellar mass at redshift of zero, we can obtain its mass assembly history assuming that it follows the evolution of SFMS.  After obtaining its stellar mass as a function of redshift, we can also obtain its size at any redshift assuming it follows the mass-size relation as calibrated in the present work (see Equation \ref{eq.model}).   In this case, we obtain the typical evolution trajectory on the mass-size plane for individual star-forming galaxies.


Specifically, we adopt the evolution of the sSFR from \cite{Lilly_2016}, follows the formula: 
\begin{equation} \label{eq:sfms}
  {\rm sSFR}(M_*,z) = \frac{0.07}{1-R} \times  (\frac{M_*}{3\times 10^{10} \mathrm{M}_{\odot}})^{-0.2}\times (1+z)^2 \ {\rm Gyr^{-1}},
\end{equation}
where $R$ indicates the fraction of mass formed in new stars that is subsequently returned to the interstellar medium through winds and supernova explosions. 
This slightly sub-linear evolving SFMS  in Equation \ref{eq:sfms} is broadly consistent with many observational results \citep[e.g.][]{2009ApJ...698L.116P, 2013ApJ...763..129S, 2014ApJS..214...15S}.  Throughout this work, the stellar mass is defined as the mass of living stars. In this case, the $1-R$ term can be omitted in the calculation. 


In Figure \ref{Fig.8}, we present the typical size evolution locus for individual galaxies. In the left panel, dashed lines show the mass-size relations at $1.0 \ \mu \mathrm{m}$ of different redshifts taken from our model.  The solid curves show the typical evolutionary paths for galaxies of different stellar masses, color-coding with the sSFR.  As can be seen following a given locus, the locus of star-forming galaxy on the mass-size plane is slightly steeper than the mass-size relation at early epoch, and becomes steeper and steeper with increasing stellar mass at late epoch. 
In addition, \cite{2018MNRAS.474.3976G} find that in TNG simulation individual galaxies grow gently on mass-size diagram at low mass, but this growth becomes increasingly steep as their masses increase. This is qualitatively in agreement with our results.
\cite{2015ApJ...813...23V} also investigated the size growth track of individual galaxies based on HST and Keck.
Our results show a similar slope for low-mass galaxies compared to \cite{2015ApJ...813...23V}, while exhibit a steeper slope for more massive galaxies.

At fixed stellar mass, we find the slope of size growth trajectory on mass-size plane (the left panel of Figure \ref{Fig.8}) shows very weak or no dependence on redshift, while this slope clearly shows a strong dependence on stellar mass at any given redshift.  As discussed before, we propose that the size ratio $R_{0.45\mu \mathrm{m}} / R_{1.0\mu \mathrm{m}}$  can be a tracer of the moving trend of individual galaxies on mass-size plane.  Here we examine this by directly plotting the correlation between the slope of locus and the observed $R_{0.45\mu \mathrm{m}} / R_{1.0\mu \mathrm{m}}$ for several specific points on the mass-size plane, marked with blue dots in the left panel of Figure \ref{Fig.8}.  For a specific point, the value of $R_{0.45\mu \mathrm{m}} / R_{1.0\mu \mathrm{m}}$ is calculated as the median value of JADES galaxies that are selected with stellar mass located within $\pm$0.4 dex and redshift within $\pm$0.5 from the point.  The result in shown in Figure \ref{Fig.ratio_slope}. 

As can be clearly seen, the $R_{0.45\mu \mathrm{m}} / R_{1.0\mu \mathrm{m}}$ from observed galaxies indeed strongly correlates with the slope of locus.  We emphasize that the slope of locus is only derived from the combination of the evolution of mass-size relation and SFMS.  However, the results of Figure \ref{Fig.8} and \ref{Fig.ratio_slope} show good consistency with the result of size variation with rest-frame wavelength in Figure \ref{Fig.2}.  This good agreement strongly suggests that the variation of size on wavelength  contains important information on size growth of galaxies on short timescales,
indicating that dust attenuation does not dominate the size variation with wavelength.

We note that our analysis does not suggest that galaxies grow faster over time.  The slope of locus on mass-size plane reflects the size growth rate with increasing a given fraction of stellar mass, rather than the absolute size growth rate in unit of kpc/Gyr.  We therefore directly show the typical size growth of individual galaxies in dashed lines in the right panel of Figure \ref{Fig.8}.  As expected, the sizes of galaxies increase much faster than the size-redshift relation (obtained in Equation \ref{eq.model}) when including the growth of stellar mass.  Interestingly, the size growth of galaxies appears to be exponential as a function of redshift.  
It is important to emphasize that the size growth history of individual galaxies may be complex. Our results reveal a typical size growth trajectory for a population of galaxies with specified stellar mass and redshift; however, the behavior of any single galaxy may not strictly adhere to this trajectory, but rather follows it statistically.

There is an assumption in deriving the size growth of individual galaxies that the size growth of star-forming galaxies are primarily driven by the in-situ star formation.  This is not unreasonable since the smooth accretion plays a dominated role than the mergers in the lifetime of galaxies \citep[e.g.][]{2002ApJ...571....1M, 2008A&ARv..15..189S, 2010ApJ...718.1001B, 2012A&A...544A..68L, 2015MNRAS.449...49R}. However, the effect of mergers on the size evolution of star-forming galaxies is not included in our novel approach.  We also point out that star formation quenching may also play a role in the mass-size evolution, since quenched galaxies drop out from the star-forming population.  If quenching depends on galaxy size and compact galaxies more likely to be quenched \citep{2013ApJ...776...63F, 2016ApJ...827L..32B, Wang-18}, average sizes of star-forming galaxies will increase with redshift purely from this effect.  It remains difficult to discriminate these effects from the observation, while analyzing the simulation data in principle can quantify these effects.

\section{summary and conclusion}\label{sec:summary}

We investigate the size variation with rest-frame wavelength for star-forming galaxies based on JADES data release version 2.0.  If ignoring the dust attenuation, the size variation with wavelength can reflect the distribution of newly formed stars with respect to older stellar populations. In other words, it indicates the size growth of galaxies by in-situ star formation. After matching PSF to F444W band, we measure the half light radius of galaxies with {\sc Galfit} by modeling the light distribution with a single \sersic\ profile for each bands.   We reveal the relationship between sizes of galaxies and their mass, redshift, and rest-frame wavelength. 


Here we summarize the key findings in this work. 
\begin{enumerate}
    \item Star-forming galaxies are typically smaller at longer wavelength from UV-to-NIR with $z<3.5$, especially for more massive galaxies.  
    This is consistent with the inside-out formation scenario.  The size variation with wavelength show strong dependence on stellar mass, and show little or not dependence on redshift, specific star formation rate and galaxy environment.  
    This indicates that newly formed stars in more massive galaxies are distributed more extensively compared to old stellar populations. 

    
    \item 
    We model size as a function of both mass and redshift simultaneously, obtaining $R_{\rm e} \propto M_*^{0.23} (1+z)^{-1.04}$ at wavelength of 0.45 ${\mu \mathrm{m}}$, and $R_{\rm e} \propto M_*^{0.20} (1+z)^{-1.08}$ at 1.0 ${\mu \mathrm{m}}$. 
    Galaxy size evolves faster at $0.45 \ \mu \mathrm{m}$ than $1.0 \ \mu \mathrm{m}$, which is consistent with the fact that massive galaxies have higher size ratio of $0.45 \ \mu \mathrm{m}$ and $1.0 \ \mu \mathrm{m}$. 
    This simplified model with only 3 free parameters is able to describe the relation between mass, size and redshift very well, and obtains consistent result with previous works.

    \item 
    We obtain the typical size growth track of individual galaxies on mass-size diagram based on the combination of our size evolution model and the star formation main sequence. 
    The growth trend strongly depends on stellar mass, with more massive galaxies having steeper slope of locus on mass-size diagram.  This independent approach results in  consistency results with what we find with size variation in Section \ref{sec:3.1}.   
    In addition, the moving trend on mass-size plane strongly correlates with the size ratio between 0.45 ${\mu \mathrm{m}}$ and 1.0 ${\mu \mathrm{m}}$. 
    
\end{enumerate}


Our result suggests that the size growth of star-forming galaxies is a self-regulated process primarily governed by stellar mass, rather than the star formation status or environment.  In addition, we find at $z>4$, galaxies tend to be smaller at shorter wavelength than at longer wavelength, which may be the signature of wet compaction.  Interestingly, \cite{tripodi2024spatially} found similar result using the spatially resolved emission lines of 63 low-mass galaxies at $4<z<10$ from JADES. 
Due to the limitation of the sample size, we only study the first order result of size variation with wavelength, and do not consider detailed dependence on galaxies properties  with controlling the stellar mass.  The growing JWST dataset with combing different surveys will enable more detailed analysis, and examine the possibly compaction process at $z>4$.   

The coherence between the moving trend on mass-size plane and the size ratio $R_{0.45\mu \mathrm{m}} / R_{1.0\mu \mathrm{m}}$, supports that the size variation with wavelength indeed reflects the information on size growth of galaxies on short timescales, rather than the pure effect of dust attenuation. However, we agree that the dust attenuation could still play a subdominant role in interpreting the results, depending on its detailed 2-dimensional distributions. In the future, we plan to take the dust effect into account with SED modelling, to extract the SFR and stellar mass surface density of galaxies.  This will provide the instantaneous size growth rate in unit of kpc/Gyr for individual galaxies, potentially uncovering the co-evolution of galaxies in both size growth and mass assembly. 

\section{acknowledgements}\label{sec:acknowledgements}
We thank the referee for their insightful comments, which have significantly improved the quality of the paper. 
CJ and EW thank Zhiyuan Ji for providing the mPSF of JADES field, and thank for Dandan Xu for useful discussion. 
EW thanks support of the National Science Foundation of China (Nos. 12473008) and the Start-up Fund of the University of Science and Technology of China (No. KY2030000200). Y.R. acknowledges supports from the CAS Pioneer Hundred Talents Program (Category B), as well as the USTC Research Funds of the Double First-Class Initiative (grant No. YD2030002013). HL is supported by the National Key R\&D Program of China No. 2023YFB3002502, the National Natural Science Foundation of China under No. 12373006 and the China Manned Space Project.
HYW is supported by the National Natural Science Foundation of China(Nos. 12192224) and CAS Project for Young Scientists in Basic Research, Grant No. YSBR-062.
The authors gratefully acknowledge the support of Cyrus Chun Ying Tang Foundations.

\appendix

\section{sample distribution}
\begin{figure}[ht]
    \centering
    \includegraphics[scale=0.6]{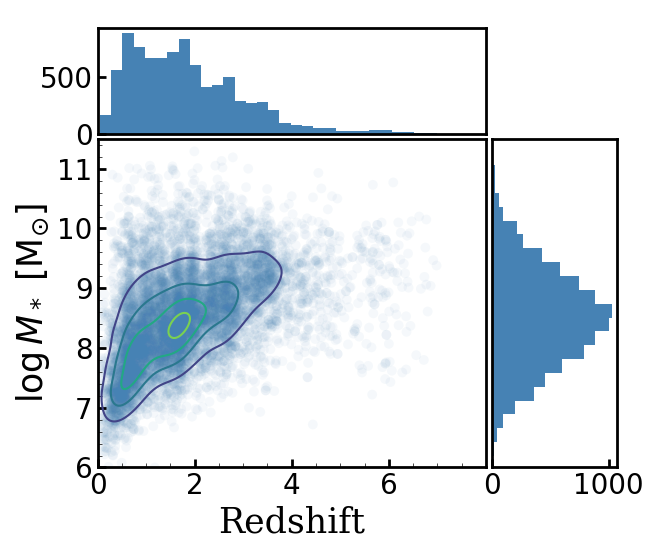}
    \caption{The distribution of our sample. The top panel shows stellar mass distribution. The right panel shows redshift distribution. The central panel represent the 2D distribution of them.}
    \label{Fig.distr}
\end{figure}

Figure \ref{Fig.distr} shows the distributions of the sample used in this work on the stellar mass and redshfit diagram. The top panel illustrates the distribution of stellar mass, while the right panel represents the distribution of redshift. As seen, the majority of galaxies are concentrated within a stellar mass range of $10^7 \mathrm{M_{\odot}}$ to $10^{10} \mathrm{M_{\odot}}$, at redshifts around 0 to 4. Although the sample is incomplete at high redshift and low stellar mass, this is not expected to significantly affect our results, since we mainly investigate the size variations on the mass-redshift diagram.  

\section{filters and corresponding rest-frame wavelength}

\begin{figure}[ht]
    \centering
    \includegraphics[scale=0.6]{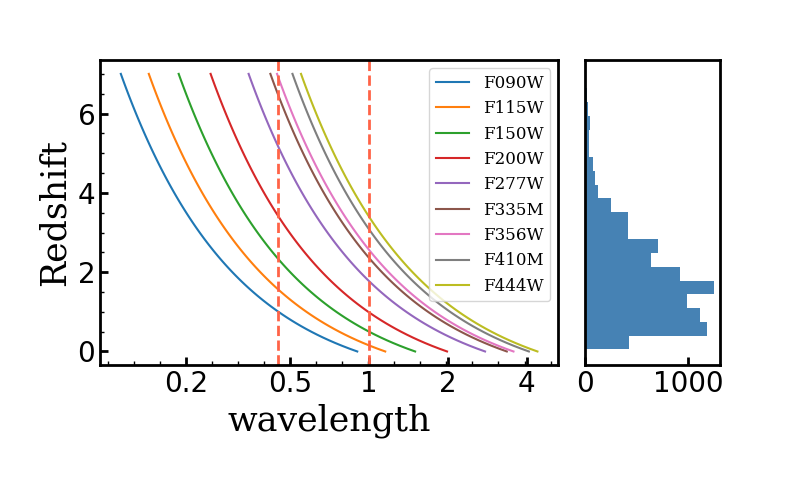}
    \caption{Filters we used and redshift distribution. It shows rest-frame wavelength at different redshift for each filters. Lines in different color represent the pivot wavelength of these bands. The target wavelength of our interpolation is indicated by the red dashed line.}
    \label{Fig.filter}
\end{figure}

We measure the sizes of galaxies in the nine bands (F090W, F115W, F150W, F200W, F277W, F335M, F356W, F410M, and F444W), which cover an observed wavelength range of approximately $0.9 - 4.4 \mu \mathrm{m}$. The rest-frame wavelengths corresponding to different redshifts for these nine bands are depicted in the Figure \ref{Fig.filter}. In addition, we show the redshift distribution of the sample in the right panel of Figure \ref{Fig.filter}.
We denote the two wavelengths $0.45 \mu \mathrm{m}$ and $1.0 \mu \mathrm{m}$, which are investigated in the main text, in red dashed lines. We note that for galaxies with $z < \sim1$ or $z > \sim3.5$, we could not find two bands at rest-frame that encompass the desired two wavelengths. Therefore, we do not include the galaxies with $z < 1$ and $z > 3.5$ during the fitting process.

\bibliography{ref}

\end{document}